\documentclass[a4paper,fleqn,usenatbib,useAMS]{mnras}

\usepackage{multicol}
\usepackage{hyperref}
\usepackage{float}

\usepackage{graphicx}	% Including figure files
\usepackage{amsmath}	% Advanced maths commands
\usepackage{amssymb}	% Extra maths symbols
\usepackage{multicol}        % Multi-column entries in tables
\usepackage{bm}		% Bold maths symbols, including upright Greek
\usepackage{pdflscape}	% Landscape pages

\newcommand{\hublow}{$\bar{H}_0 \pm \sigma_{H_0} = 68 \pm 2.8$ km s$^{-1}$ Mpc$^{-1}$}
\newcommand{\hubhigh}{$\bar{H}_0 \pm \sigma_{H_0} = 73.24 \pm 1.74$ km s$^{-1}$ Mpc$^{-1}$}
\newcommand{\lcdm}{$\Lambda$CDM }
\newcommand{\pcdm}{$\phi$CDM }

\usepackage[T1]{fontenc}
\usepackage{ae,aecompl}

\usepackage{newtxtext,newtxmath}
\title[Constraints on dark energy and spatial curvature]{Constraints on dark energy dynamics and spatial curvature from Hubble parameter and baryon acoustic oscillation data}

\author[]{Joseph Ryan$^{1}$\thanks{E-mail:jwryan@phys.ksu.edu},
    Sanket Doshi$^{2, 1}$\thanks{E-mail: sanketdoshik2@iitb.ac.in}, and
    Bharat Ratra$^{1}$\thanks{E-mail:ratra@phys.ksu.edu}
\\
$^1$Department of Physics, Kansas State University, 116 Cardwell Hall, Manhattan, KS 66506, USA \\
$^2$Department of Physics, Indian Institute of Technology Bombay, Mumbai 400076, India \\
}

\date{Last updated 20XX YYY ZZ; in original form 20XX YYY ZZ}

\pubyear{2018}

\begin{document}
\label{firstpage}
\pagerange{\pageref{firstpage}--\pageref{lastpage}}
\maketitle

\begin{abstract}
We use all available baryon acoustic oscillation distance measurements and Hubble parameter data to constrain the cosmological constant $\Lambda$, dynamical dark energy, and spatial curvature in simple cosmological models. We find that the consensus spatially flat $\Lambda$CDM model provides a reasonable fit to the data, but depending on the Hubble constant prior and cosmological model, it can be a little more than 1$\sigma$ away from the best-fit model, which can favor mild dark energy dynamics or non-flat spatial hypersurfaces.
\end{abstract}


%%%%%%%%%%%%%%%%%%%%%%%%%%%%%%%%%%%%%%%%%%%%%%%%%%

%%%%%%%%%%%%%%%%% BODY OF PAPER %%%%%%%%%%%%%%%%%%

% The MNRAS class isn't designed to include a table of contents, but for this document one is useful.
% I therefore have to do some kludging to make it work without masses of blank space.

\defcitealias{4}{Planck Collaboration (2016)}

\section{Introduction}
\label{sec. 1}

It is widely accepted that the universe is undergoing accelerated expansion today. The consensus cosmological model, $\Lambda$CDM, posits that this acceleration is driven by the spatially homogeneous, constant dark energy density $\rho_{\Lambda}$ of the cosmological constant $\Lambda$ \citep{65}. For reviews of the accelerated cosmological expansion and of the $\Lambda$CDM model, see \cite{66}, \cite{63}, \cite{64}, and \cite{62}. In this model, cold dark matter (CDM) is the second largest contributor to the current energy budget and, with non-relativistic baryonic matter, powered the decelerating cosmological expansion at earlier times. 

The consensus $\Lambda$CDM model assumes flat spatial hypersurfaces, but observations don't rule out mildly curved spatial hypersurfaces; observations also do not rule out the possibility that the dark energy density varies slowly with time. In this paper we examine, in addition to the general (not necessarily spatially flat) $\Lambda$CDM model, the XCDM parametrization of dynamical dark energy, and the $\phi$CDM model in which a scalar field $\phi$ is the dynamical dark energy.\footnote{While cosmic microwave background (CMB) anisotropy data provide the most restrictive constraints on cosmological parameters, many other measurements have been used to constrain the XCDM parametrization and the $\phi$CDM model \citetext{see, e.g., \citealp{24}, \citealp{31}, \citealp{25}, \citealp{21}, \citealp{20}, \citealp{23}, \citealp{19}, \citealp{26}, \citealp{27, 29, 30, 28}, \citealp{18}, \citealp{22}, \citealp{32}, \citealp{91}, \citealp{92}}.} In the XCDM and $\phi$CDM cases we allow for both vanishing and non-vanishing spatial curvature. Details of the three models we study are summarized in Sec. \ref{sec. 2}, and more information can be found in \cite{5}.

\cite{36} have recently shown that, in the spatially flat case, the Planck 2015 CMB anisotropy data from \citetalias{4} (and some baryon acoustic oscillation distance measurements) weakly favor the XCDM parametrization and the $\phi$CDM model of dynamical dark energy over the $\Lambda$CDM consensus model. The XCDM case results have been confirmed by \cite{38} for a much bigger compilation of cosmological data, including most available Type Ia supernova apparent magnitude observations, BAO distance measurements, growth factor data, and Hubble parameter observations.\footnote{For earlier indications favoring dynamical dark energy over the $\Lambda$CDM consensus model, based on smaller compilations of data, see \cite{39}, \cite{34}, \cite{40}, \cite{43}, \cite{28}, \cite{26}, \cite{29}, \cite{42}, \cite{27}, \cite{41}, \cite{30}, \cite{22}, \cite{33}, and \cite{35}. However, more recent analyses, based on bigger compilations of data, do not support the significant evidence for dynamical dark energy indicated in some of the earlier analyses \citep{36, 38}.} Also, spatially flat XCDM and $\phi$CDM both reduce the tension between CMB temperature anisotropy and weak gravitational lensing estimates of $\sigma_8$, the rms fractional energy density inhomogeneity averaged over $8\hspace{1mm}h^{-1}$Mpc radius spheres, where $h$ is the Hubble constant in units of 100 km s$^{-1}$ Mpc$^{-1}$ \citep{36, 38}.
\defcitealias{4}{Planck Collaboration 2016}

In non-flat models nonzero spatial curvature provides an additional length scale which invalidates usage of the power-law power spectrum for energy density inhomogeneities in the non-flat case (as was assumed in the analysis of non-flat models in \citetalias{4}). Non-flat inflation models \citep{46, 47, 55} provide the only known physically-consistent mechanism for generating energy density inhomogeneities in the non-flat case; the resulting open and closed model power spectra are not power laws \citep{57, 58, 56}. \defcitealias{4}{Planck Collaboration (2016)}Using these power spectra, \cite{50} have found that the Planck 2015 CMB anisotropy data in combination with a few BAO distance measurements no longer rule out the non-flat $\Lambda$CDM case (unlike the earlier \citetalias{4} analyses based on the incorrect assumption of a power-law power spectrum in the non-flat model).\footnote{Currently available non-CMB measurements do not significantly constrain spatial curvature \citep{45, 44, 61, 48, 12, 59, 54, 60, 49}.} \cite{53} confirmed these results for a bigger compilation of cosmological data, and similar conclusions hold in the non-flat dynamical dark energy XCDM and $\phi$CDM cases \citep{51, 52, 38}.

Additionally, the non-flat models provide a better fit to the observed low multipole CMB temperature anisotropy power spectrum, and do better at reconciling the CMB anisotropy and weak lensing constraints on $\sigma_8$, but do a worse job at fitting the observed large multipole CMB anisotropy temperature power spectrum \citep{50, 51, 52, 38, 53}. Given the non-standard normalization of the Planck 2015 CMB anisotropy likelihood and that the flat and non-flat $\Lambda$CDM models are not nested, it is not possible to compute the relative goodness of fit between the flat and non-flat $\Lambda$CDM models quantitatively, although qualitatively the flat $\Lambda$CDM model provides a better fit to the current data \citep{50, 51, 52, 38, 53}.

In the analyses discussed above, the Planck 2015 CMB anisotropy data played the major role. Those authors found consistency between cosmological constraints derived using the CMB anisotropy data in combination with various non-CMB data sets. CMB anisotropy data are sensitive to the behavior of cosmological spatial inhomogeneities. Here we derive constraints on similar models from a combination of all available Hubble parameter data as well as all available radial and transverse BAO data.\footnote{The $H(z)$ and radial BAO data provide a unique measure of the cosmological expansion rate over a wide redshift range, up to almost $z = 2.4$, well past the cosmological deceleration-acceleration transition redshift. These data show evidence for this transition and can be used to measure the redshift of the transition \citep{67, 74, 88, 68, 12, 60, 89, 93}.} Unlike the CMB anisotropy data, the $H(z)$ and these BAO data are not sensitive to the behavior of cosmological spatial inhomogeneities.

The models that we study, and the methods we use to analyze these data, are the same as those presented in \cite{12, 45}, and we also use some of the same $H(z)$ and baryon acoustic oscillation measurements. We differ from those studies by now using all currently available $H(z)$ and baryon acoustic oscillation data.

The constraints we derive here are consistent with, but weaker than, those of the papers cited above; this provides a necessary and useful consistency test of those results. In particular, we find that the consensus flat $\Lambda$CDM model is a reasonable fit, in most cases, to the BAO and $H(z)$ data we study here. However, depending somewhat on the Hubble constant prior we use, consensus flat $\Lambda$CDM can be 1$\sigma$ away from the best-fit parameter values in some cases, which can favor mild dark energy dynamics or non-flat spatial hypersurfaces.

In Sec. \ref{sec. 2} we provide a short summary of the models we studied. Sec. \ref{sec. 3} presents the data that we used, and in Sec. \ref{sec. 4} we describe the methods by which we analyzed these data. Sec. \ref{sec. 5} describes the results of our analyses, and our conclusions are given in Sec. \ref{sec. 6}.

\section{Models}
\label{sec. 2}
The models we examine in this paper are characterized by their expansion rate as a function of redshift $z$,
\begin{equation}
E(z) = \frac{H(z)}{H_0}.
\end{equation}
Here $H(z)$ is the Hubble parameter and $H_0 \equiv H(0)$ is the Hubble constant.

In the $\Lambda$CDM model dark energy is a constant vacuum energy density with negative pressure, equivalent to an ideal fluid with equation of state parameter
\begin{equation}
w = \frac{p_b}{\rho_b} = -1.
\end{equation}
Here $p_b$ and $\rho_b$ are the homogeneous parts of the pressure and energy density, respectively. The expansion rate can be written in terms of the density parameters
\begin{equation}
E(z) = \sqrt{\Omega_{m0}\left(1 + z\right)^3 + \left(1 - \Omega_{m0} - \Omega_{\Lambda}\right)\left(1 + z\right)^2 + \Omega_{\Lambda}},
\end{equation}
where $\Omega_{m0}$ is the current value of the non-relativistic matter density parameter, $\Omega_{\Lambda}$ is the cosmological constant energy density parameter, and $\Omega_{k0} = 1 - \Omega_{m0} - \Omega_{\Lambda}$ (which is nonzero in general) is the current value of the spatial curvature energy density parameter. Here, and in the other models we study, we ignore the contributions from CMB photons and neutrinos, which are very small at the redshifts of the data we use, so the $\Lambda$CDM model is characterized by two parameters: $p = \left(\Omega_{m0}, \Omega_{\Lambda}\right)$.

In the XCDM parametrization of dark energy, $w = w_X$ where $w_X$ is a negative constant (in general $w_X \neq -1$). Hence
\begin{equation}
E(z) = \sqrt{\Omega_{m0}\left(1 + z\right)^3 + \Omega_{k0}\left(1 + z\right)^2 + \Omega_{X0}\left(1 + z\right)^{3\left(1 + w_X\right)}},
\end{equation}
where $\Omega_{X0}$ is the current value of the dark energy density. In contrast to $\Lambda$CDM, the dark energy density parameter $\Omega_{X0}\left(1 + z\right)^{3\left(1 + w_X\right)}$ varies with time.\footnote{In the XCDM parametrization, the energy density and pressure of the dark energy fluid, $\rho_{Xb}(t)$ and $p_{Xb}(t)$, are space-independent functions of time. When $\rho_{Xb}(t)$ is negative, this is an inconsistent parametrization that is rendered consistent by assuming a constant speed of acoustic inhomogeneities (typically $c_{sX} = 1$). The BAO and $H(z)$ data we consider only constrain the spatially homogeneous part of the cosmological models.} If, however, $w_X = -1$, then XCDM reduces to $\Lambda$CDM, with $\Omega_{X0} = \Omega_{\Lambda}$. In general, the XCDM parametrization has three free parameters: $p = \left(\Omega_{m0}, \Omega_{k0}, w_X\right).$ We shall also consider spatially flat XCDM, with $p = \left(\Omega_{m0}, w_X\right)$.

The $\phi$CDM model \citep{6, 14, 5, 17} provides a simple, physically consistent description of dynamical dark energy. In this model, the dark energy is a scalar field $\phi$ with a potential energy density given by
\begin{equation}
V(\phi) = \frac{1}{2}\kappa m_p^2 \phi^{-\alpha}.
\end{equation}
Here $\alpha > 0$, $m_p^2 \equiv G^{-1}$, $G$ is the gravitational constant, and
\begin{equation}
\kappa = \frac{8}{3}\left(\frac{\alpha + 4}{\alpha + 2}\right)\left[\frac{2}{3}\alpha(\alpha + 2)\right]^{-\alpha/2}.
\end{equation}
The spatially homogeneous part of the scalar field obeys
\begin{equation}
\ddot{\phi} + 3\frac{\dot{a}}{a}\dot{\phi} - \frac{1}{2}\kappa \alpha m_p^2\phi^{-\alpha-1} = 0,
\end{equation}
where $a = a(t)$ is the scale factor, and an overdot denotes differentiation with respect to time. This, together with the first Friedmann equation,
\begin{equation} \label{eq. 8}
\left(\frac{\dot{a}}{a}\right)^2 = \frac{8\pi G}{3}\left(\rho_m + \rho_{\phi}\right) - \frac{k}{a^2},
\end{equation}
and
\begin{equation} 
    \rho_{\phi} = \frac{1}{2}\dot{\phi}^2 + V(\phi),
\end{equation}
determines the dynamics of the field. In eq. (\ref{eq. 8}), $\rho_m$ is the non-relativistic matter density, $\rho_{\phi}$ is the scalar field energy density, and $k = 0, +1, -1$ for flat, closed, and open spatial hypersurfaces, respectively.

The dark energy equation of state parameter of $\phi$CDM is
\begin{equation}
w_{\phi}  = \frac{p_{\phi}}{\rho_{\phi}} = \frac{\frac{1}{2}\dot{\phi}^2 - V(\phi)}{\frac{1}{2}\dot{\phi}^2 + V(\phi)},
\end{equation}
which, unlike in the $\Lambda$CDM and XCDM parametrizations, changes with time. The expansion rate in the $\phi$CDM model is
\begin{equation}
E(z) = \sqrt{\Omega_{m0}\left(1 + z\right)^3 + \Omega_{k0}\left(1 + z\right)^2 + \Omega_{\phi}(z,\alpha)},
\end{equation}
where
\begin{equation}
\Omega_{\phi}(z, \alpha) \equiv \frac{8 \pi G\rho_{\phi}}{3 H_0^2}.
\end{equation}
In contrast to $\Omega_{X}$, $\Omega_{\phi}$ is not an explicit function of a power of $\left(1 + z\right)$; it must be determined numerically.

In general, the $\phi$CDM model has three free parameters: $p = \left(\Omega_{m0}, \Omega_{k0}, \alpha\right)$. We also consider spatially flat $\phi$CDM with $p = \left(\Omega_{m0}, \alpha\right)$.

\section{Data}
\label{sec. 3}

\begin{table}
\caption{BAO data. $D_M \left(r_{s,{\rm fid}}/r_s\right)$ and $D_V \left(r_{s,{\rm fid}}/r_s\right)$ have units of Mpc, while $H(z)\left(r_s/r_{s,{\rm fid}}\right)$ has units of ${\rm km}\hspace{1mm}{\rm s}^{-1}{\rm Mpc}^{-1}$ and $r_s$ has units of Mpc.}
\label{table 1}
\begin{tabular}{ccccc}
\hline
$z$ & Measurement & Value & $\sigma$ & Ref.\\
\hline
$0.38$ & $D_M\left(r_{s,{\rm fid}}/r_s\right)$ & 1518 & 22 & \cite{1}\\
\hline
$0.51$ & $D_M\left(r_{s,{\rm fid}}/r_s\right)$ & 1977 & 27 & \cite{1}\\
\hline
$0.61$ & $D_M\left(r_{s,{\rm fid}}/r_s\right)$ & 2283 & 32 & \cite{1}\\
\hline
$0.38$ & $H(z)\left(r_s/r_{s,{\rm fid}}\right)$ & 81.5 & 1.9 & \cite{1}\\
\hline
$0.51$ & $H(z)\left(r_s/r_{s,{\rm fid}}\right)$ & 90.4 & 1.9 & \cite{1}\\
\hline
$0.61$ & $H(z)\left(r_s/r_{s,{\rm fid}}\right)$ & 97.3 & 2.1 & \cite{1}\\
\hline
$0.106$ & $r_s/D_V$ & 0.336 & 0.015 & \cite{10}\\
\hline
$0.15$ & $D_V\left(r_{s,{\rm fid}}/r_s\right)$ & $664$ & $25$ & \cite{2}\\
\hline
$1.52$ & $D_V\left(r_{s,{\rm fid}}/r_s\right)$ & $3855$ & $170$ & \cite{3}\\
\hline
$2.33$ & $\frac{\left(D_H\right)^{0.7} \left(D_{M}\right)^{0.3}}{r_s}$ & 13.94 & 0.35 & \cite{9}\\
\hline
$2.36$ & $c/\left(r_s H(z)\right)$ & 9.0 & 0.3 & \cite{11}\\
\hline
\end{tabular}
\end{table}

\begin{table}
\caption{$H(z)$ data. $H(z)$ and $\sigma_H$ have units of ${\rm km}\hspace{1mm}{\rm s}^{-1}{\rm Mpc}^{-1}$.}
\label{table 2}
\begin{tabular}{cccc}
\hline
$z$ & $H(z)$ & $\sigma_{H}$ & Ref.\\
\hline
0.07 & 69 & 19.6 & \cite{73}\\
\hline
0.09 & 69 & 12 & \cite{69}\\
\hline
0.12	& 68.6 & 26.2 & \cite{73}\\
\hline
0.17	& 83 & 8 & \cite{69}\\
\hline
0.179	& 75 & 4 & \cite{70}\\
\hline
0.199	& 75 & 5 & \cite{70}\\
\hline
0.20	& 72.9 & 29.6 & \cite{73}\\
\hline
0.27	& 77 & 14 & \cite{69}\\
\hline
0.28	& 88.8 & 36.6 & \cite{73}\\
\hline
0.352	& 83 & 14 & \cite{70}\\
\hline
0.3802 & 83 & 13.5 & \cite{68}\\
\hline
0.4	& 95 & 17 & \cite{69}\\
\hline
0.4004 & 77 & 10.2 & \cite{68}\\
\hline
0.4247 & 87.1 & 11.2 & \cite{68}\\
\hline
0.4497 & 92.8 & 12.9 & \cite{68}\\
\hline
0.47 & 89 & 50 & \cite{15}\\
\hline
0.4783 & 80.9 & 9 & \cite{68}\\
\hline
0.48	& 97 & 62 & \cite{71}\\
\hline
0.593	& 104	& 13 & \cite{70}\\
\hline
0.68	& 92 & 8 & \cite{70}\\
\hline
0.781	& 105	& 12 & \cite{70}\\
\hline
0.875	& 125	& 17 & \cite{70}\\
\hline
0.88	& 90 & 40 & \cite{71}\\
\hline
0.90	& 117 & 23 & \cite{69}\\
\hline
1.037	& 154	& 20 & \cite{70}\\
\hline
1.3	& 168	& 17 & \cite{69}\\
\hline
1.363	& 160	& 33.6 & \cite{72}\\
\hline
1.43	& 177	& 18 & \cite{69}\\
\hline
1.53	& 140	& 14 & \cite{69}\\
\hline
1.75	& 202	& 40 & \cite{69}\\
\hline
1.965	& 186.5 & 50.4 & \cite{72}\\
\hline
\end{tabular}
\end{table}

BAO provide observers with a ``standard ruler" which can be used to measure cosmological distances (see \citealp{75} for a review). These distances can be computed in a given cosmological model, so measurements of them can be used to constrain the parameters of the model in question. The BAO distance measurements we use are listed in Table \ref{table 1}.

The transverse co-moving distance is
\begin{equation}
    D_M(z) = 
    \begin{cases}
    D_C & \text{if}\ \Omega_{k0} = 0,\\
    \vspace{1mm}
    \frac{c}{H_0\sqrt{\Omega_{k0}}}{\rm sinh}\left[\sqrt{\Omega_{k0}}\frac{D_C H_0}{c}\right] & \text{if}\ \Omega_{k0} > 0, \\
    \vspace{1mm}
    \frac{c}{H_0\sqrt{|\Omega_{k0}|}}{\rm sin}\left[\sqrt{|\Omega_{k0}|}\frac{D_C H_0}{c}\right] & \text{if}\ \Omega_{k0} < 0,
    \end{cases}
\end{equation}
where
\begin{equation}
    D_H = \frac{c}{H(z)},
\end{equation}
\begin{equation}
    D_C = \frac{c}{H_0}\int^z_0 \frac{dz'}{E(z')},
\end{equation}
and the volume-averaged angular diameter distance is
\defcitealias{4}{Planck Collaboration 2016}
\begin{equation}
    D_V(z) = \left[\frac{cz}{H_0}\frac{D^2_M(z)}{E(z)}\right]^{1/3}
\end{equation}
\citep{7, 5}. All of the measurements in Table \ref{table 1} are scaled by the size of the sound horizon at the drag epoch ($r_{\rm s}$). This quantity is \citetext{see \citealp{8} for a derivation}:
\begin{equation}
    r_{\rm s} = \frac{2}{3k_{\rm eq}}\sqrt{\frac{6}{R_{\rm eq}}}{\rm ln}\left[\frac{\sqrt{1 + R_d} + \sqrt{R_d + R_{\rm eq}}}{1 + \sqrt{R_{\rm eq}}}\right]
\end{equation}
where $R_d \equiv R(z_d)$ and $R_{\rm eq} \equiv R(z_{\rm eq})$ are the values of $R$, the ratio of the baryon to photon momentum density,
\begin{equation}
    R = \frac{3\rho_b}{4\rho_{\gamma}}
\end{equation}
at the drag epoch and matter-radiation equality epoch, respectively. Here $k_{\rm eq}$ is the scale of the particle horizon at the matter-radiation equality epoch, and $\rho_b$ and $\rho_{\gamma}$ are the baryon and photon mass densities. In our analyses, where appropriate, the original data listed in Table \ref{table 1} have been rescaled to a fiducial sound horizon $r_{\rm s, fid} = 147.60$ Mpc \citetext{from Table 4, column 3, of \citetalias{4}}. This fiducial sound horizon was determined by using the $\Lambda$CDM model, so its value is model dependent, though not to a significant degree (as can be seen by comparing the computed $r_s$ of the \citetalias{4} baseline model to that measured using the spatially open \lcdm and flat XCDM parametrization of \citetalias{4}).

In Table \ref{table 2} we list 31 $H(z)$ measurements determined using the cosmic chronometric technique, which are the same as the cosmic chronometric $H(z)$ data used in \cite{60} \citetext{see e.g. \citealp{70} for a discussion of cosmic chronometers}. With this method, the Hubble rate as a function of redshift is determined by using
\begin{equation}
    H(z) = -\frac{1}{\left(1 + z\right)}\frac{dz}{dt}.
\end{equation}
Although this determination of $H(z)$ does not depend on a cosmological model, it does depend on the quality of the measurement of $dz/dt$, which requires an accurate determination of the age-redshift relation for a given chronometer. See \cite{70} and \cite{72} for discussions of the strengths and weaknesses of this method. While their approach requires accurate knowledge of the star formation history and metallicity of massive, passively evolving early galaxies, and although the two different techniques they use give slightly different values, they also point out that the measurement of $H(z)$ from this method is relatively insensitive to changes in the chosen stellar population synthesis model.

\defcitealias{4}{Planck Collaboration (2016)}

\begin{figure*}
\begin{multicols}{3}
    \includegraphics[width=\linewidth]{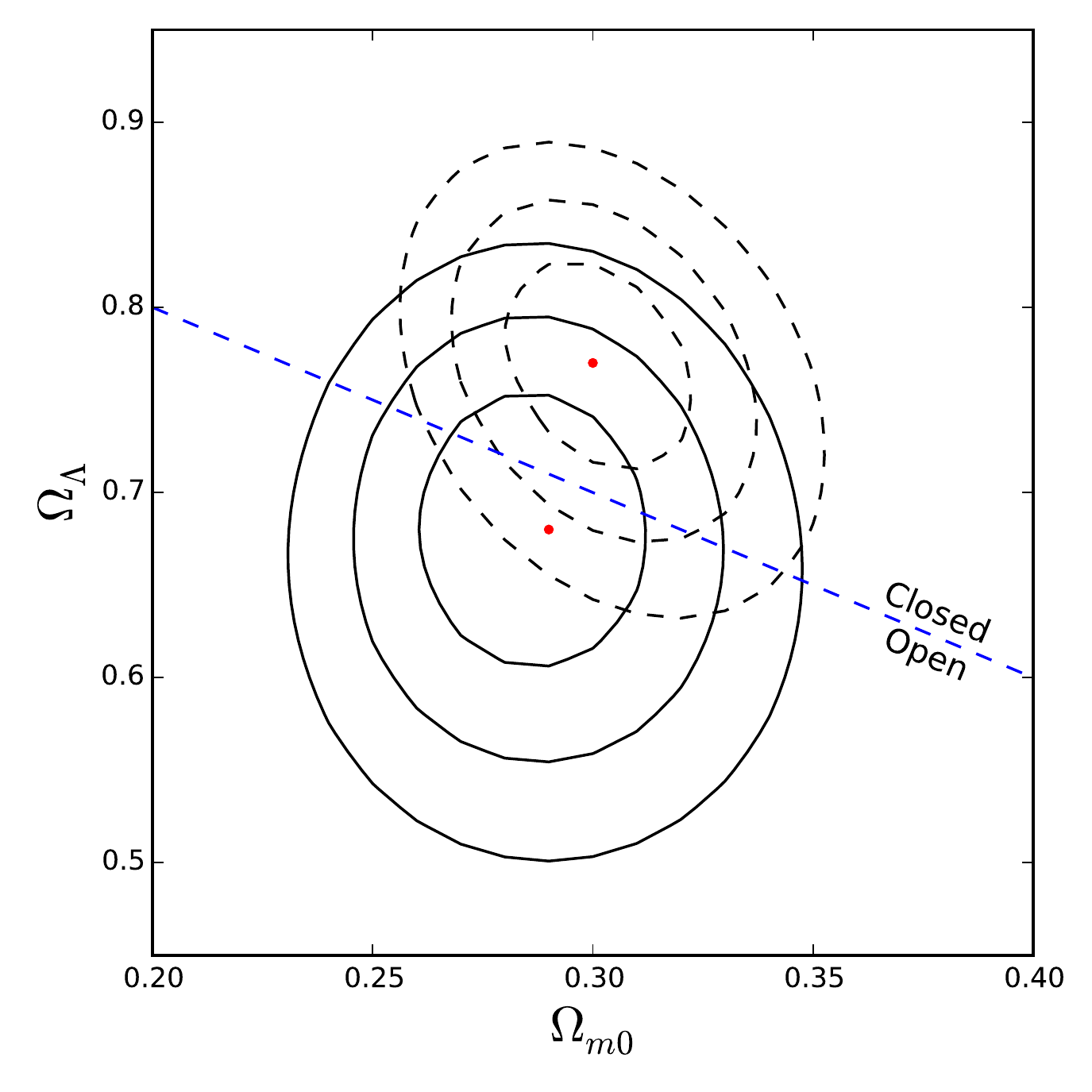}\par 
    \includegraphics[width=\linewidth]{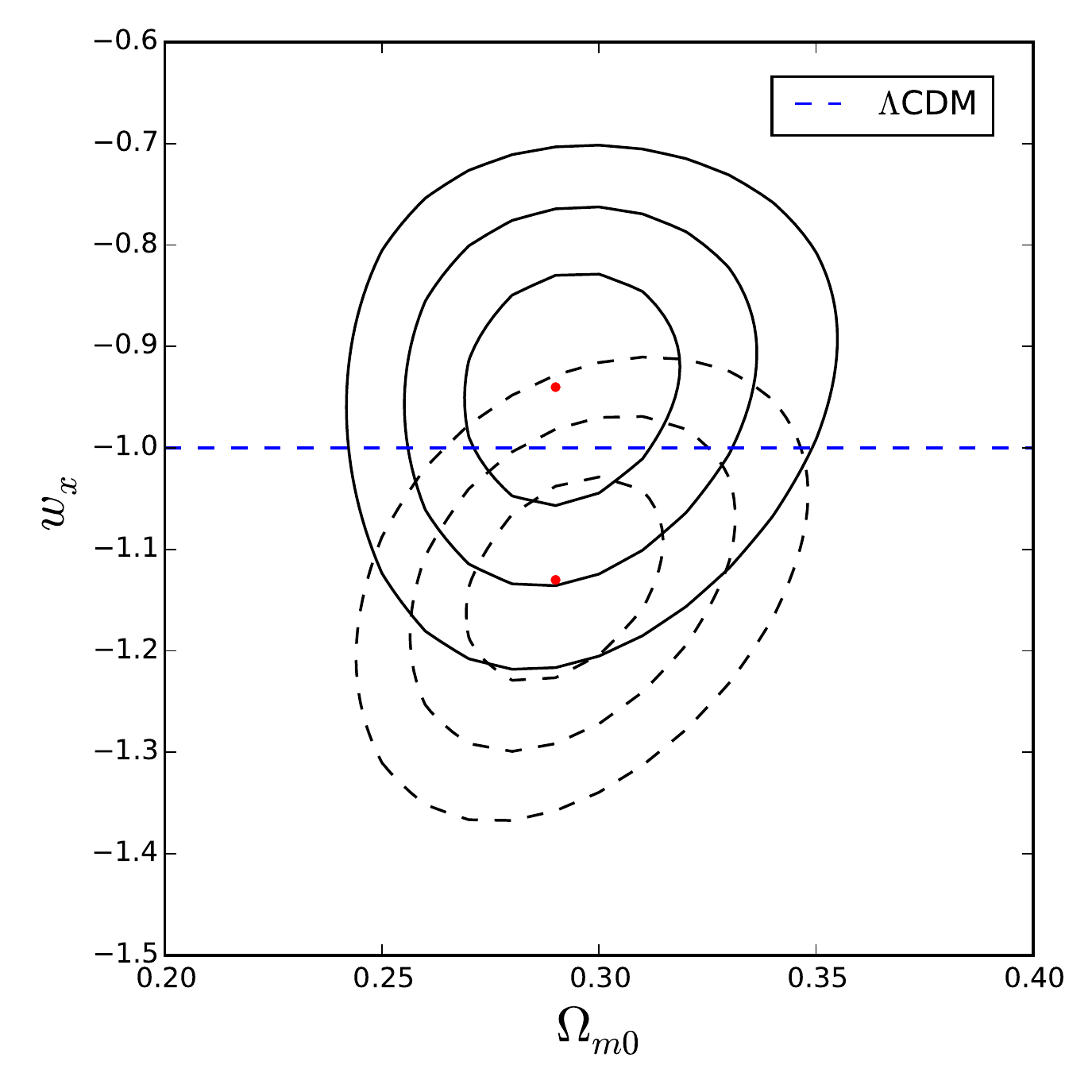}\par
    \includegraphics[width=\linewidth]{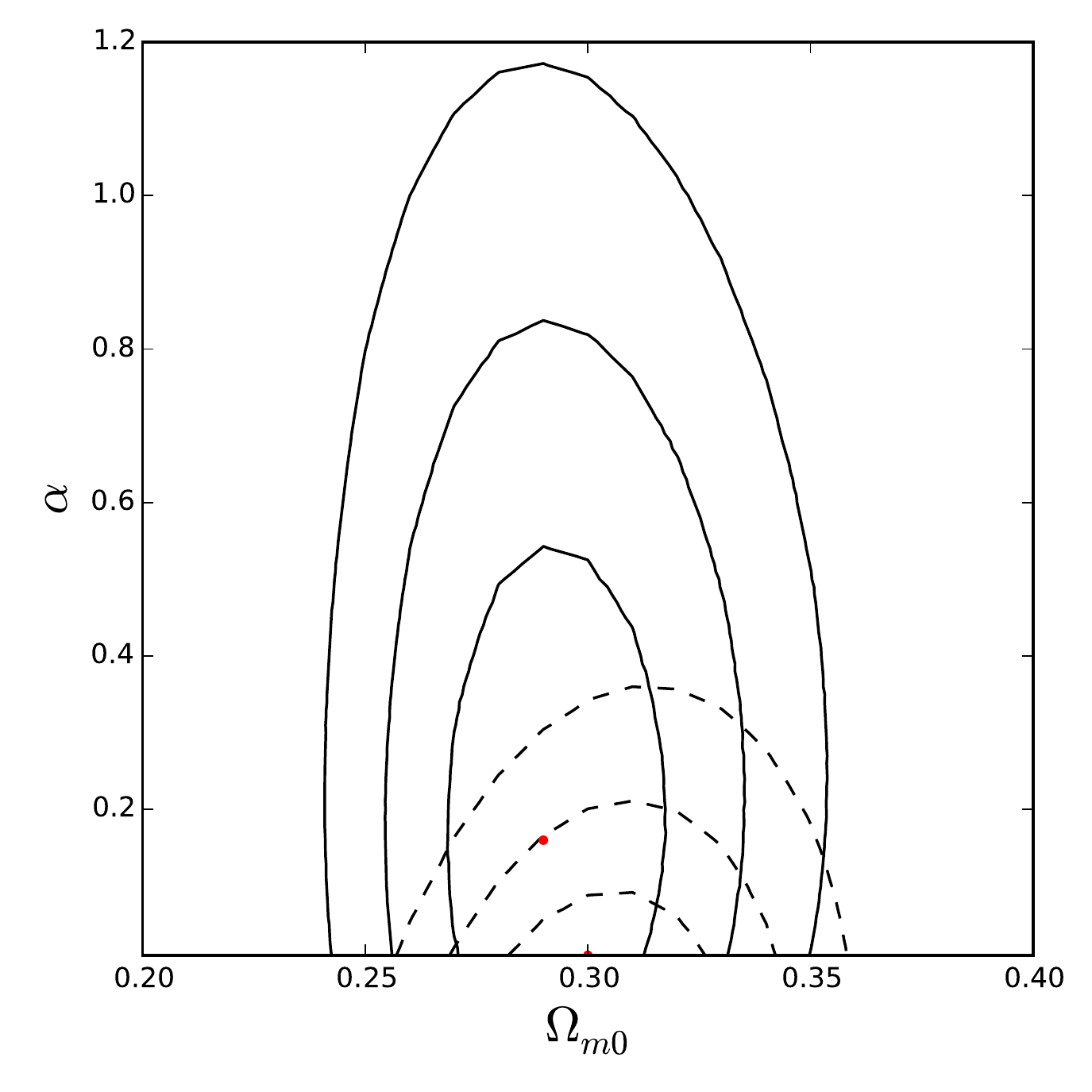}\par
\end{multicols}
\caption{Confidence contours for 2-parameter models. Solid (dashed) 1, 2, and 3$\sigma$ contours correspond to $\bar{H}_0 \pm \sigma_{H_0} = 68 \pm 2.8 \hspace{1mm}(73.24 \pm 1.74)$ km s$^{-1}$ Mpc$^{-1}$ prior, and the red dots indicate the location of the best-fit point in each prior case. Left: $\Lambda$CDM. The blue dashed line indicates the spatially flat $\Lambda$CDM model; points above (below) the line correspond to models with closed (open) spatial hypersurfaces. Center: flat XCDM. The blue dashed line (for which $w_{\rm X} = -1$) demarcates the flat $\Lambda$CDM case. Right: flat $\phi$CDM. The horizontal $\alpha = 0$ axis corresponds to the flat $\Lambda$CDM model. Color online.}
\label{fig. 1}
\end{figure*}

\begin{figure*}
\begin{multicols}{3}
    \includegraphics[width=\linewidth]{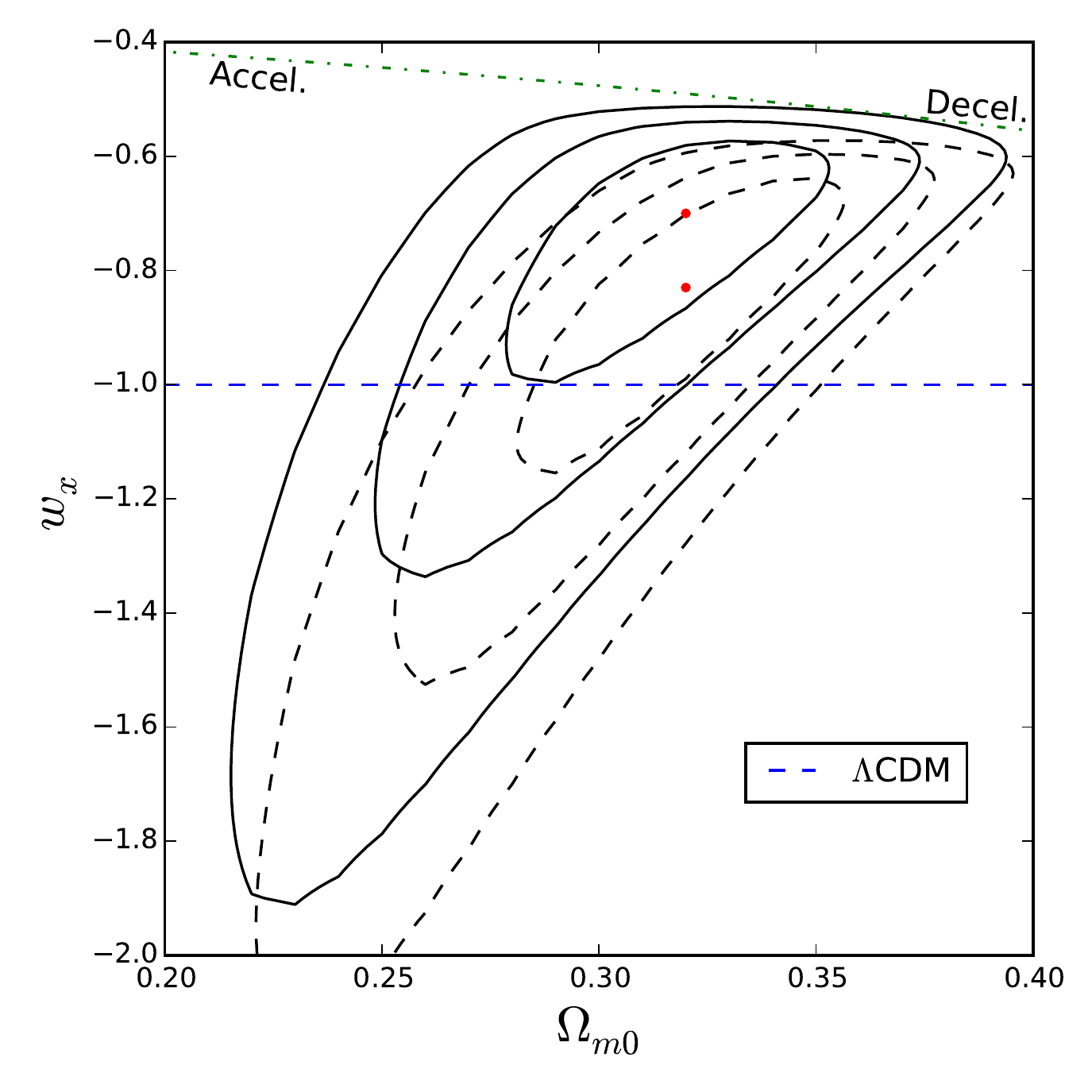}\par 
    \includegraphics[width=\linewidth]{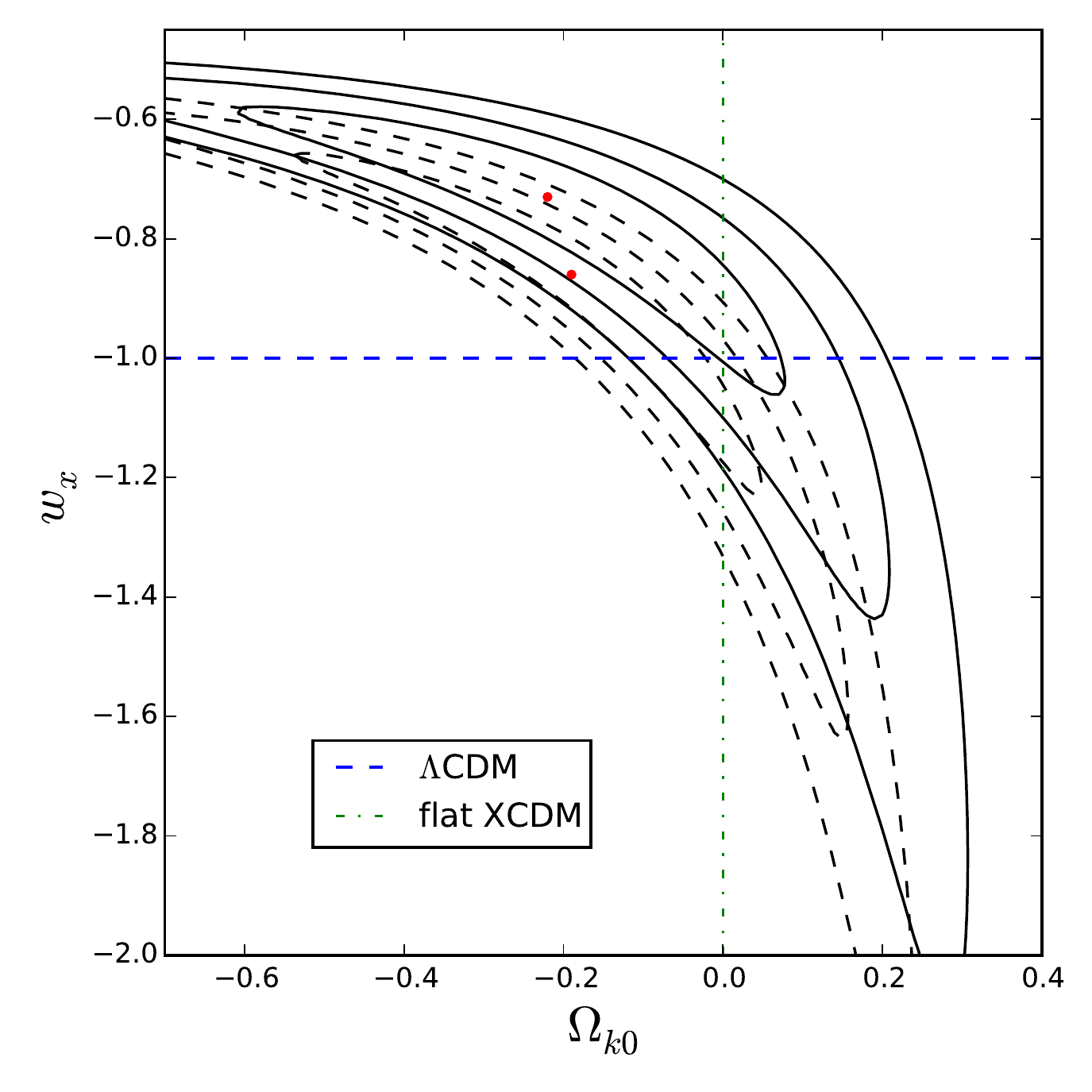}\par
    \includegraphics[width=\linewidth]{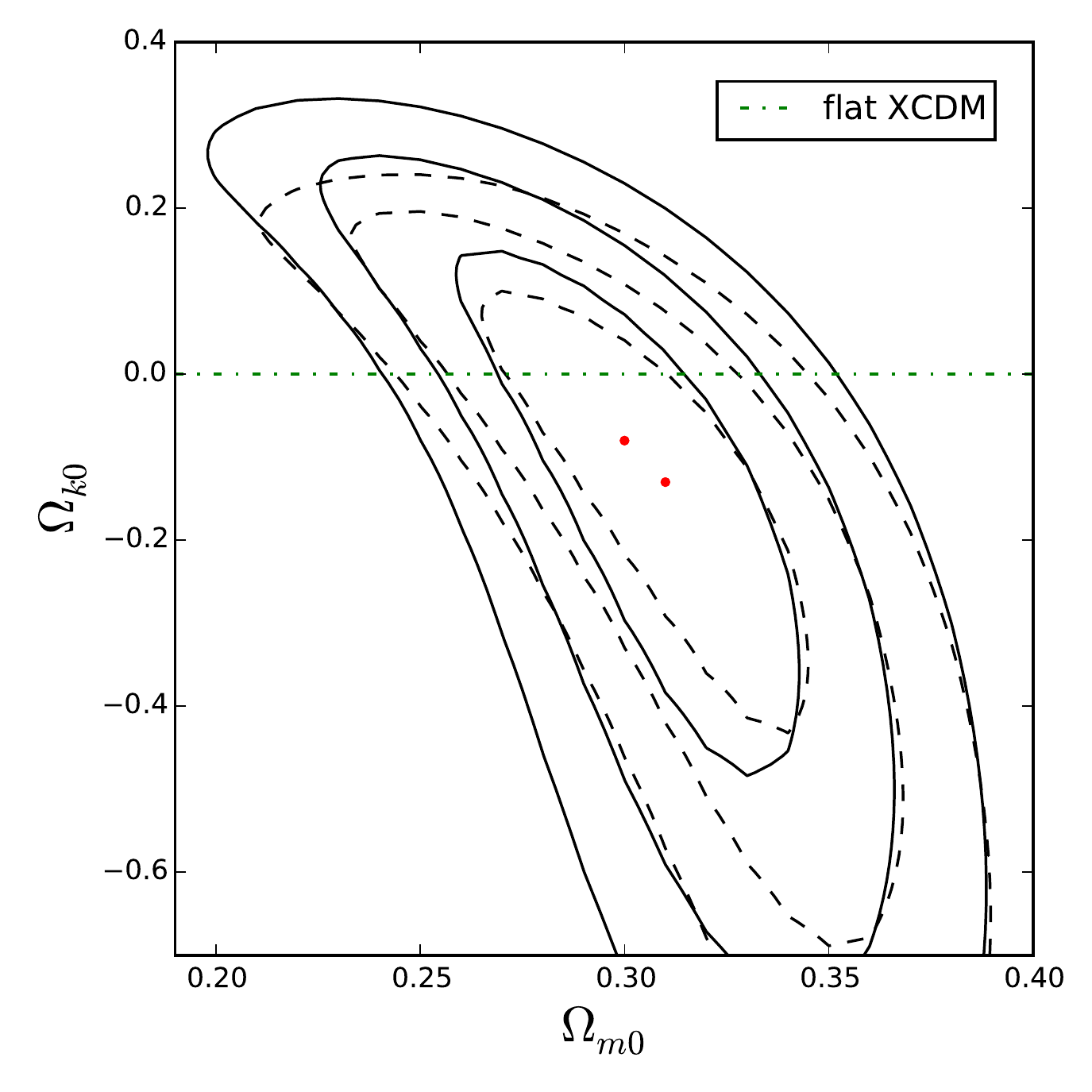}\par
\end{multicols}
\caption{Confidence contours for 3-parameter XCDM. Solid (dashed) 1, 2, and 3$\sigma$ contours correspond to $\bar{H}_0 \pm \sigma_{H_0} = 68 \pm 2.8 \hspace{1mm}(73.24 \pm 1.74)$ km s$^{-1}$ Mpc$^{-1}$ prior, and the red dots indicate the location of the best-fit point in each prior case. Left: $\Omega_{k0}$ marginalized. The blue dashed line indicates the $\Lambda$CDM model. Points above (below) the green dot-dashed curve near the top of the panel correspond to models with late-time (decelerating) accelerating expansion. Center: $\Omega_{m0}$ marginalized. The horizontal blue dashed line (for which $w_{\rm X} = -1$) demarcates the $\Lambda$CDM case, and the vertical green dot-dashed line demarcates the spatially flat XCDM case. Right: $w_X$ marginalized. The horizontal green dot-dashed line indicates the spatially flat XCDM case. Color online.}
\label{fig. 2}
\end{figure*}

\section{Methods}
\label{sec. 4}
\defcitealias{4}{Planck Collaboration 2016}

	To determine the values of the best-fit parameters, we minimized
\begin{equation}
\chi^2(p) \equiv -2 ln \mathcal{L}(p),
\end{equation}
where $\mathcal{L}$ is the likelihood function and $p$ is the set of parameters of the model under consideration. If the likelihood function $\mathcal{L}(p, \nu)$ depends on an uninteresting nuisance parameter $\nu$ with a probability distribution $\pi(\nu)$, we marginalize the likelihood function by integrating $\mathcal{L}(p, \nu)$ over $\nu$
\begin{equation}
\mathcal{L}(p) = \int \mathcal{L}(p, \nu) \pi(\nu) d\nu.
\end{equation}
In our $H(z)$ analyses $H_0$ is a nuisance parameter. We assumed a Gaussian distribution for $H_0$
\begin{equation}
    \pi\left(H_0\right) = \frac{1}{\sqrt{2\pi\sigma^2_{H_0}}}{\rm exp}\left[\frac{-\left(H_0 - \bar{H}_0\right)^2}{2\sigma^2_{H_0}}\right]
\end{equation}
and marginalized over it. We considered two cases: $\bar{H}_0 \pm \sigma_{H_0} = 68 \pm 2.8$ km s$^{-1}$ Mpc$^{-1}$ and $\bar{H}_0 \pm \sigma_{H_0} = 73.24 \pm 1.74$ km s$^{-1}$ Mpc$^{-1}$.\footnote{The lower value, $68 \pm 2.8$ km s$^{-1}$ Mpc$^{-1}$ is the most recent median statistics estimate of the Hubble constant \citep{13}. It is consistent with earlier median statistics estimates \citep{80, 76}. It is also consistent with many other recent measurements of $H_0$ \citetext{\citetalias{4}; \citealp{48}; \citealp{77}; \citealp{85}; \citealp{82}; \citealp{81}; \citealp{87}; \citealp{60}; \citealp{38}; \citealp{93}}. The higher value, $73.24 \pm 1.74$ km s$^{-1}$ Mpc$^{-1}$, comes from a local expansion rate estimate \citep{83}. Other local expansion rate estimates find slightly lower $H_0$'s with larger error bars \citep{84, 86, 78, 79}.}

Most of the data we analyzed are uncorrelated, however six of the data points \citetext{those from \citealp{1}}, are correlated. For uncorrelated data points,
\begin{equation} \label{eq. 23}
\chi^2(p) = \sum^{N}_{i = 1} \frac{[A_{{\rm th}}(p; z_i) - A_{{\rm obs}}(z_i)]^2}{\sigma_i^2},
\end{equation}
where $A_{{\rm th}}(p; z_i)$ are the model predictions at redshifts $z$, and $A_{{\rm obs}}(z_i)$ and $\sigma_i$ are the central values and error bars of the measurements listed in Table \ref{table 2} and the last five lines of Table \ref{table 1}. The correlated data (the first six entries in Table \ref{table 1}) require
\begin{equation} \label{eq. 24}
\chi^2(p) = \left[\vec{A}_{{\rm th}}(p) - \vec{A}_{{\rm obs}}\right]^{T} \mathcal{C}^{-1} \left[\vec{A}_{{\rm th}}(p) - \vec{A}_{{\rm obs}}\right]
\end{equation}
where $\mathcal{C}^{-1}$ is the inverse of the covariance matrix
\begin{equation}
    \mathcal{C} = 
\begin{bmatrix}
    484.0       & 9.530 & 295.2 & 4.669 & 140.2 & 2.402 \\
    9.530       & 3.610 & 7.880 & 1.759 & 5.983 & 0.9205 \\
    295.2 & 7.880 & 729.0 & 11.93 & 442.4 & 6.866 \\
    4.669 & 1.759 & 11.93 & 3.610 & 9.552 & 2.174 \\
    140.2 & 5.983 & 442.4 & 9.552 & 1024 & 16.18 \\
    2.402 & 0.9205 & 6.866 & 2.174 & 16.18 & 4.410 \\
\end{bmatrix}
\end{equation}
\citep{1}. $\vec{A}_{\rm obs}$ (in eq. \ref{eq. 24}) are the measurements in the first six lines of Table \ref{table 1}.
	
In addition to $\chi^2$, we also used the Bayes Information Criterion
\begin{equation}
{\rm BIC} \equiv \chi^2_{{\rm min}} + k {\rm ln} N
\end{equation}
and the Akaike Information Criterion
\begin{equation}
{\rm AIC} \equiv \chi^2_{{\rm min}} + 2k
\end{equation}
\citep{90}. In these equations $\chi^2_{\rm min}$ is the minimum value of $\chi^2$, $k$ is the number of parameters of the given model, and $N$ is the number of data points. BIC and AIC provide means to compare models with different numbers of parameters; they penalize models with a higher $k$ in favor of those with a lower $k$, in effect enforcing Occam's Razor in the model selection process.

To determine the confidence intervals $r_n$ on the 1d best-fit parameters, we computed one-sided limits $r^{\pm}_n$ by using
\begin{equation} \label{eq. 28}
    \frac{\int^{r^{\pm}_n}_{\bar{p}} \mathcal{L}(p)dp}{\int^{\pm \infty}_{\bar{p}} \mathcal{L}(p)dp} = \sigma_n,
\end{equation}
where $\bar{p}$ is the point at which $\mathcal{L}(p)$ has its maximimum value, such that $n = 1, 2$ and $\sigma_1 = 0.6827$, $\sigma_2 = 0.9545$. Because the one-dimensional likelihood function is not guaranteed to be symmetric about $\bar{p}$, we compute the upper and lower confidence intervals separately. In the $\Lambda$CDM model, for example, the 1-sigma confidence intervals on $\Omega_{m0}$ are computed by first integrating the likelihood function $\mathcal{L}(\Omega_{m0}, \Omega_{\Lambda})$ over $\Omega_{\Lambda}$ to obtain a marginalized likelihood function that only depends on $\Omega_{m0}$,
\begin{equation}
    \int^1_0 \mathcal{L}(\Omega_{m0}, \Omega_{\Lambda})d\Omega_{\Lambda} = \mathcal{L}(\Omega_{m0}),
\end{equation}
and then inserting this marginalized likelihood function into eq. (\ref{eq. 28}).

The ranges over which we marginalized the parameters of the $\Lambda$CDM model were $0 \leq \Omega_{\Lambda} \leq 1$ and $0.01 \leq \Omega_{m0} \leq 1$. For the spatially flat XCDM parametrization, we used $-2 \leq w_X \leq 0$ and $0.01 \leq \Omega_{m0} \leq 1$, and for the spatially flat $\phi$CDM model we used $0.01 \leq \alpha \leq 5$ and $0.01 \leq \Omega_{m0} \leq 1$. For 3-parameter XCDM, we used $-0.7 \leq \Omega_{k0} \leq 0.7$, $0.01 \leq \Omega_{m0} \leq 1$, and $-2.00 \leq w_X \leq 0$. For the 3-parameter $\phi$CDM model we considered $-0.5 \leq \Omega_{k0} \leq 0.5$, $0.01 \leq \Omega_{m0} \leq 1$, and $0.01 \leq \alpha \leq 5$. \footnote{$\Omega_{m0}, \alpha = 0.01$ were excluded because our codes ran into difficulties at those points.} 

We analyzed the data with two independent Python codes, written by S.D. and J.R., that produced almost identical results in the 2-parameter cases and the 3-parameter XCDM parametrization, and results that agreed to within 1\% in the 3-parameter $\phi$CDM case.

\section{Results}
\label{sec. 5}
\begin{figure*}
\begin{multicols}{3}
    \includegraphics[width=\linewidth]{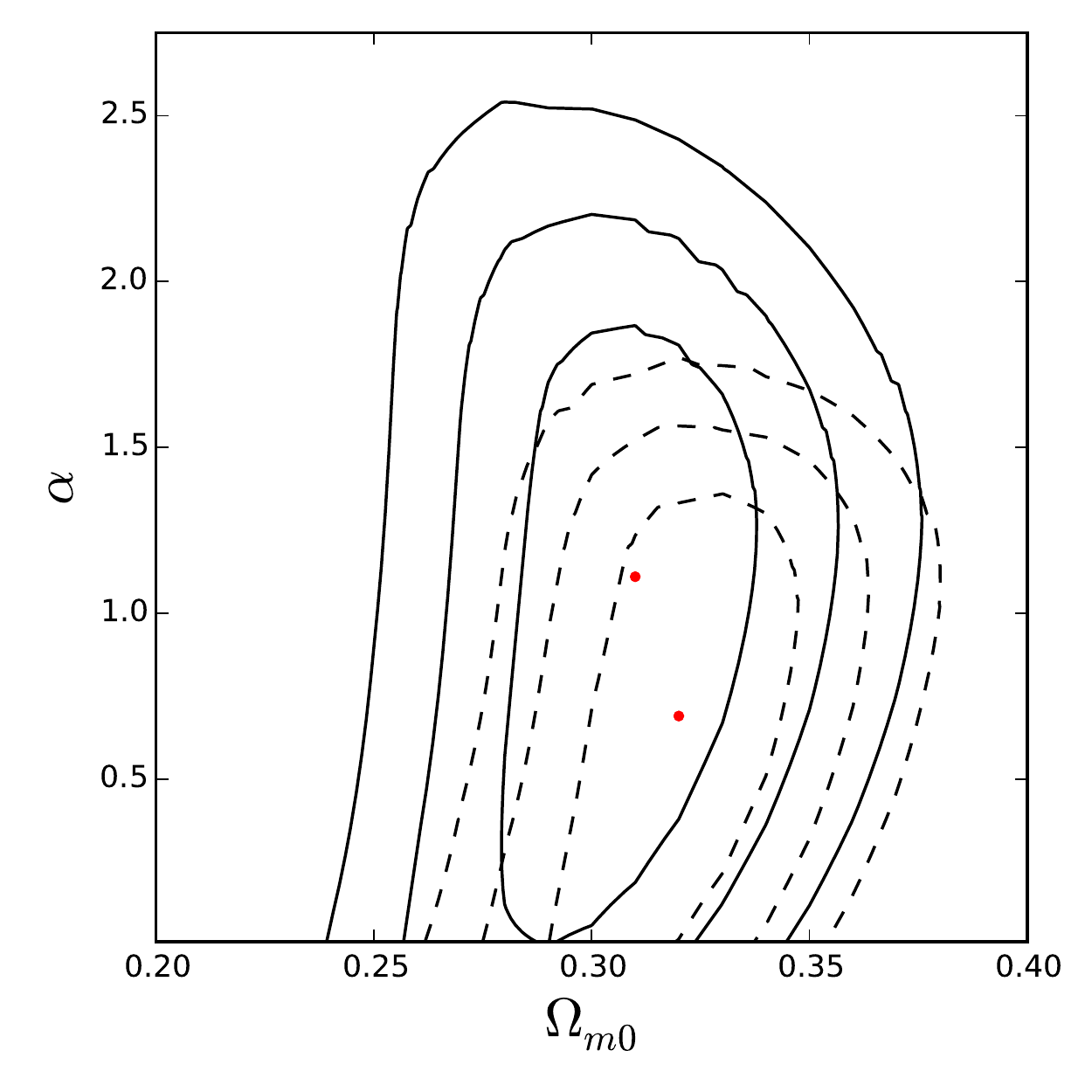}\par 
    \includegraphics[width=\linewidth]{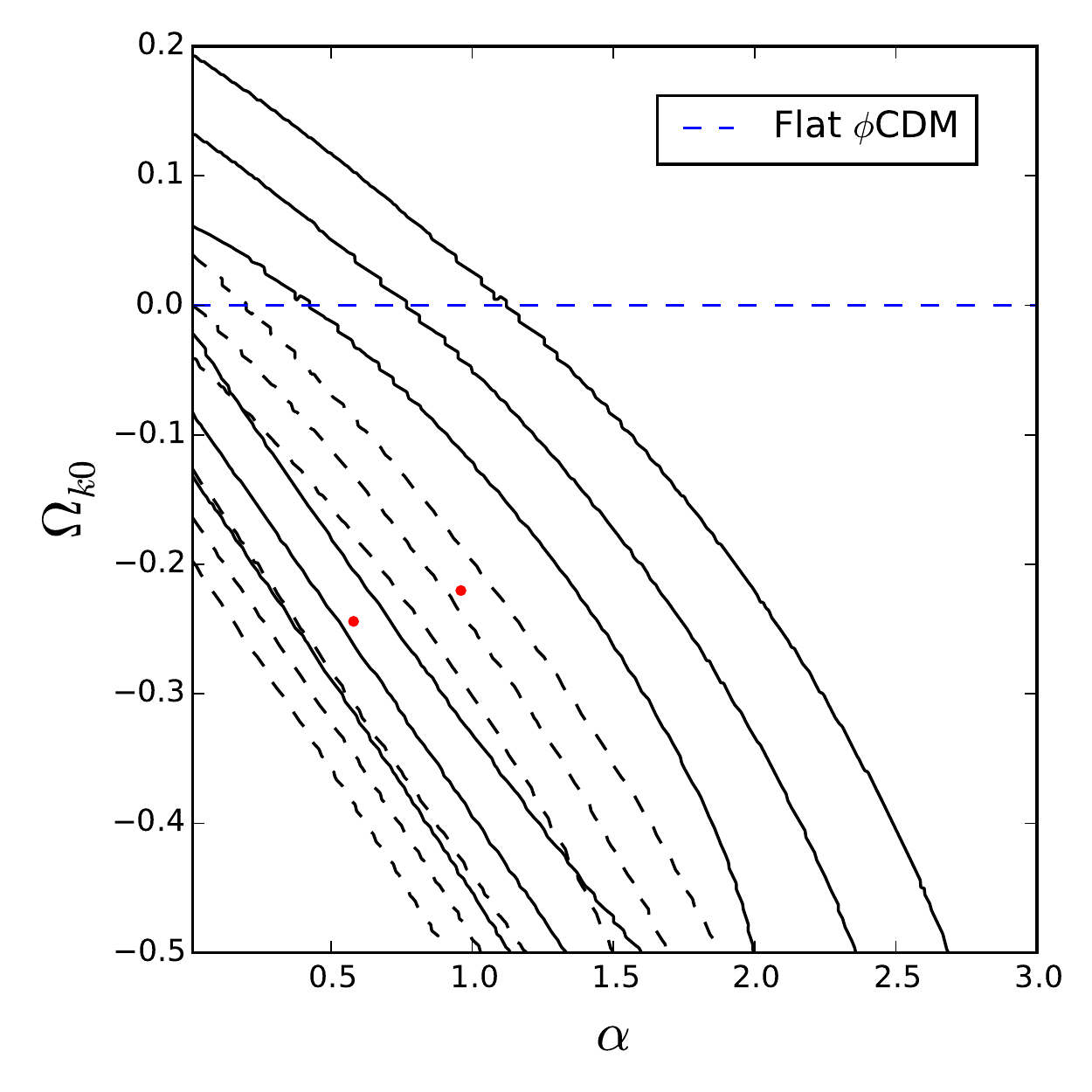}\par
    \includegraphics[width=\linewidth]{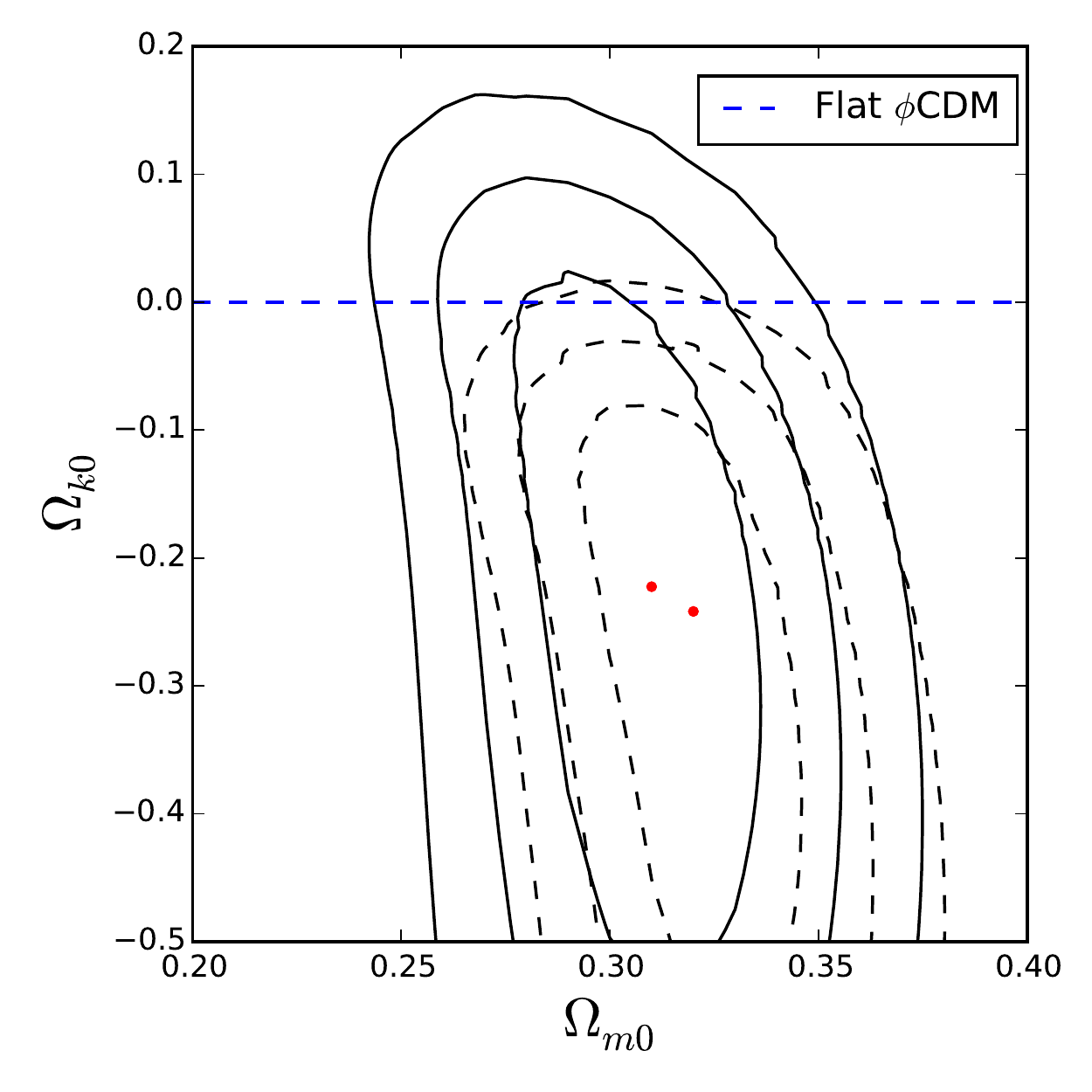}\par
\end{multicols}
\caption{Confidence contours for 3-parameter $\phi$CDM. Solid (dashed) 1, 2, and 3$\sigma$ contours correspond to $\bar{H}_0 \pm \sigma_{H_0} = 68 \pm 2.8 \hspace{1mm}(73.24 \pm 1.74)$ km s$^{-1}$ Mpc$^{-1}$ prior, and the red dots indicate the location of the best-fit point in each prior case. Left: $\Omega_{k0}$ marginalized. The horizontal $\alpha = 0$ axis corresponds to the $\Lambda$CDM model. Center: $\Omega_{m0}$ marginalized. The vertical $\alpha = 0$ axis corresponds to the $\Lambda$CDM model and the horizontal blue dashed line here and in the next panel correspond to the spatially flat $\phi$CDM case. Right: $\alpha$ marginalized. Color online.}
\label{fig. 3}
\end{figure*}

The confidence contours for the models we considered are shown in Figs. \ref{fig. 1}, \ref{fig. 2}, and \ref{fig. 3}. The solid black contours indicate the $\bar{H}_0 = 68 \pm 2.8$ km s$^{-1}$ Mpc$^{-1}$ prior constraints, the dashed black contours indicate the $H_0 = 73.24 \pm 1.74$ km s$^{-1}$ Mpc$^{-1}$ prior constraints, and the red dots indicate the best-fit point in each prior case. Our results for the parameter values of the unmarginalized and marginalized cases are collected in Tables \ref{table 3}-\ref{table 6}, along with their $\chi^2$, AIC, and BIC values. Wherever $\Delta \chi^2$, $\Delta$AIC, and $\Delta$BIC are given, these are computed relative to the $\chi^2$, AIC, and BIC of the corresponding $\Lambda$CDM model of each prior case.

 \begin{table*}
  \caption{Best-fit values for 2-parameter models. $\Delta \chi^2$ is evaluated relative to $\chi^2$ of $\Lambda$CDM for each $H_0$ prior.}
  \label{table 3}
  \begin{tabular}{ccccccccccc}
    \hline
    $H_0$ prior $\left({\rm km}\hspace{1mm}{\rm s}^{-1}{\rm Mpc}^{-1}\right)$ & Model & $\Omega_{m0}$ & $\Omega_{\Lambda0}$ & $w_X$ & $\alpha$ & $\chi^2$ & $\Delta \chi^2$ & AIC & BIC\\
    \hline
    $68 \pm 2.8$ & $\Lambda$CDM & 0.29 & 0.68 & - & - & 25.35 & 0.00 & 29.35 & 32.83\\
     & flat XCDM & 0.29 & - & -0.94 & - & 25.04 & -0.31 & 29.04 & 32.52\\
     & flat $\phi$CDM & 0.29 & - & - & 0.16 & 25.05 & -0.30 & 29.05 & 32.53\\
    \hline
    $73.24 \pm 1.74$ & $\Lambda$CDM & 0.30 & 0.77 & - & - & 26.92 & 0.00 & 30.92 & 34.40\\
     & flat XCDM & 0.29 & - & -1.13 & - & 28.26 & 1.34 & 32.26 & 35.74\\
     & flat $\phi$CDM & 0.30 & - & - & 0.01 & 32.62 & 5.70 & 36.62 & 40.10\\
    \hline
  \end{tabular}
 \end{table*}

In the 2-parameter case, the spatially flat XCDM parametrization has the lowest value of $\chi^2$ if the prior on $H_0$ is chosen to be $\bar{H}_0 = 68 \pm 2.8$ km s$^{-1}$ Mpc$^{-1}$. If, on the other hand, the $H_0$ prior is chosen to be $\bar{H}_0 = 73.24 \pm 1.74$ km s$^{-1}$ Mpc$^{-1}$ then the spatially flat $\Lambda$CDM model has the lowest value of $\chi^2$. These models also have lower AIC and BIC values than the 3-parameter XCDM parametrization and the 3-parameter $\phi$CDM model (see Tables \ref{table 3} and \ref{table 4}). On the other hand, the 3-parameter models typically have a lower $\chi^2$ than the 2-parameter $\Lambda$CDM case. These differences, however, are not statistically significant. Focusing on the $\bar{H}_0 = 68 \pm 2.8$ km s$^{-1}$ Mpc$^{-1}$ prior case, the $\chi^2$ differences indicate that the non-flat $\phi$CDM model and non-flat XCDM parametrization provide a 1.2$\sigma$ and 1.3$\sigma$ better fit to the data, respectively, while from $\Delta$AIC we find that these two models are 79\% and 86\% as probable as the 2-parameter $\Lambda$CDM model, respectively.

In Table \ref{table 5} (\ref{table 6}), we list the 1$\sigma$ and 2$\sigma$ confidence intervals on the parameters of each of the 2-parameter (3-parameter) models. We obtained these by marginalizing the 2-parameter (3-parameter) likelihood function as described in Sec. \ref{sec. 4}. The best-fit points in these tables correspond to the maximum value of the relevant one-dimensional marginalized likelihood function. Table \ref{table 3} (\ref{table 4}) lists the corresponding two-dimensional (three-dimensional) best-fit points.

From the figures and tables, we see that the spatially flat $\Lambda$CDM model is a reasonable fit to the $H(z)$ and BAO data we use (although the flat XCDM parametrization and flat $\phi$CDM model provide slightly better fits in the \hublow case). In particular, from the figures, for the \hublow prior, flat $\Lambda$CDM is always within about 1$\sigma$ of the best-fit value. However, the \hubhigh case favors some larger deviations from flat \lcdm. For example in the middle panel of Fig. \ref{fig. 1} for the flat XCDM parametrization it favors a phantom model over flat \lcdm at a little more than 1$\sigma$, while in the center and right panels of Fig. \ref{fig. 3} for the non-flat \pcdm case it also favors a closed model at a little more than 2$\sigma$. Similar conclusions may be drawn from the parameter limits listed in Tables \ref{table 5} and \ref{table 6}.

When both dynamical dark energy and spatial curvature are present (as opposed to cases with only dynamical dark energy or only spatial curvature) it is not as easy to constrain both parameters simultaneously. This can be seen by comparing the center and right panels of Fig. \ref{fig. 1} to the left panels of Figs. \ref{fig. 2} and \ref{fig. 3}, respectively. When spatial curvature is allowed to vary, the confidence contours in the 3-parameter XCDM parametrization and the $\phi$CDM model expand along the $w_X$ and $\alpha$ axes (these are the parameters that govern the dynamics of the dark energy).

 \begin{table*}
  \caption{Best-fit values for 3-parameter models. $\Delta \chi^2$, $\Delta$AIC, and $\Delta$BIC are evaluated relative to $\chi^2$, AIC, and BIC of $\Lambda$CDM for each $H_0$ prior.}
  \label{table 4}
  \begin{tabular}{cccccccccccc}
    \hline
    $H_0$ prior $\left({\rm km}\hspace{1mm}{\rm s}^{-1}{\rm Mpc}^{-1}\right)$ & Model & $\Omega_{m0}$ & $\Omega_{k0}$ & $w_X$ & $\alpha$ & $\chi^2$ & $\Delta \chi^2$ & AIC & $\Delta$AIC & BIC & $\Delta$BIC\\
    \hline
    $68 \pm 2.8$ & XCDM & 0.31 & -0.18 & -0.76 & - & 23.65 & -1.70 & 29.65 & 0.30 & 34.86 & 2.03\\
     & $\phi$CDM & 0.31 & -0.22 & - & 0.96 & 23.82 & -1.53 & 29.82 & 0.47 & 35.03 & 2.20\\
    \hline
    $73.24 \pm 1.74$ & XCDM & 0.32 & -0.21 & -0.84 & - & 26.48 & -0.44 & 32.48 & 1.56 & 37.69 & 3.29\\
     & $\phi$CDM & 0.32 & -0.26 & - & 0.62 & 26.30 & 0.95 & 32.30 & 1.38 & 37.51 & 3.11\\
    \hline
  \end{tabular}
 \end{table*}
 
 \begin{table*}
  \caption{1$\sigma$ and 2$\sigma$ parameter intervals for 2-parameter models.}
  \label{table 5}
  \begin{tabular}{cccccc}
    \hline 
    $H_0$ prior $\left({\rm km}\hspace{1mm}{\rm s}^{-1}{\rm Mpc}^{-1}\right)$ & Model & Marginalization range & Best-fit & $1\sigma$ & $2\sigma$\\
    \hline \vspace{2pt}
     $68\pm 2.8$ & $\Lambda$CDM & $0 \leq \Omega_{\Lambda0} \leq 1$ & $\Omega_{m0} = 0.29$ & $0.27 \leq \Omega_{m0} \leq 0.31$ & $0.26 \leq \Omega_{m0} \leq 0.32$\\
     & & $0.01 \leq \Omega_{m0} \leq 1$ & $\Omega_{\Lambda0} = 0.68$ & $0.63 \leq \Omega_{\Lambda0} \leq 0.73$ & $0.58 \leq \Omega_{\Lambda0} \leq 0.77$\\
     \hline \vspace{2pt}
      & flat XCDM & $-2 \leq w_X \leq 0$ & $\Omega_{m0} = 0.29$ & $0.28 \leq \Omega_{m0} \leq 0.31$ & $0.26 \leq \Omega_{m0} \leq 0.33$\\
     &  & $0.01 \leq \Omega_{m0} \leq 1$ & $w_X = -0.94$ & $-1.02 \leq w_X \leq -0.87$ & $-1.10 \leq w_X \leq -0.80$\\ 
     \hline \vspace{2pt}
      & flat $\phi$CDM & $0.01 \leq \alpha \leq 5$ & $\Omega_{m0} = 0.29$ & $0.28 \leq \Omega_{m0} \leq 0.31$ & $0.26 \leq \Omega_{m0} \leq 0.33$\\
     &  & $0.01 \leq \Omega_{m0} \leq 1$ & $\alpha = 0.16$ & $0.06 \leq \alpha \leq 0.43$ & $0.02 \leq \alpha \leq 0.72$\\ 
    \hline \hline \vspace{2pt}
     $73.24 \pm 1.74$ & $\Lambda$CDM & $0 \leq \Omega_{\Lambda0} \leq 1$ & $\Omega_{m0} = 0.30$ & $0.29 \leq \Omega_{m0} \leq 0.32$ & $0.27 \leq \Omega_{m0} \leq 0.33$\\
     &  & $0.01 \leq \Omega_{m0} \leq 1$ & $\Omega_{\Lambda0} = 0.77$ & $0.73 \leq \Omega_{\Lambda0} \leq 0.81$ & $0.69 \leq \Omega_{\Lambda0} \leq 0.84$\\ 
     \hline \vspace{2pt}
     & flat XCDM & $-2 \leq w_X \leq 0$ & $\Omega_{m0} = 0.29$ & $0.28 \leq \Omega_{m0} \leq 0.31$ & $0.26 \leq \Omega_{m0} \leq 0.32$\\
     &  & $0.01 \leq \Omega_{m0} \leq 1$ & $w_X = -1.13$ & $-1.20 \leq w_X \leq -1.06$ & $-1.27 \leq w_X \leq -1.00$\\ 
     \hline \vspace{2pt}
     & flat $\phi$CDM & $0.01 \leq \alpha \leq 5$ & $\Omega_{m0} = 0.31$ & $0.29 \leq \Omega_{m0} \leq 0.32$ & $0.28 \leq \Omega_{m0} \leq 0.34$\\
     &  & $0.01 \leq \Omega_{m0} \leq 1$ & $\alpha = 0.01$ & $0.01 \leq \alpha \leq 0.09$ & $0.01 \leq \alpha \leq 0.20$\\ 
     \hline \vspace{2pt}
  \end{tabular}
 \end{table*}
 
  \begin{table*}
  \caption{1$\sigma$ and 2$\sigma$ parameter intervals for 3-parameter models.}
  \label{table 6}
  \begin{tabular}{cccccc}
    \hline 
    $H_0$ prior $\left({\rm km}\hspace{1mm}{\rm s}^{-1}{\rm Mpc}^{-1}\right)$ & Model & Marginalization range & Best-fit & $1\sigma$ & $2\sigma$\\
    \hline \vspace{2pt}
     $68\pm 2.8$ & XCDM & $-0.7 \leq \Omega_{k0} \leq 0.7$ & $\Omega_{m0} = 0.31$ & $0.28 \leq \Omega_{m0} \leq 0.33$ & $0.25 \leq \Omega_{m0} \leq 0.36$\\
     & & & $w_X = -0.70$ & $-0.93 \leq w_X \leq -0.62$ & $-1.27 \leq w_X \leq -0.57$\\
     \hline \vspace{2pt}
      &  & $0.01 \leq \Omega_{m0} \leq 1$ & $\Omega_{k0} = -0.11$ & $-0.36 \leq \Omega_{k0} \leq 0.06$ & $-0.59 \leq \Omega_{k0} \leq 0.19$\\
     & & & $w_X = -0.70$ & $-0.93 \leq w_X \leq -0.62$ & $-1.27 \leq w_X \leq -0.57$\\
     \hline \vspace{2pt}
     &  & $-2 \leq w_X \leq 0$ & $\Omega_{m0} = 0.31$ & $0.28\leq \Omega_{m0} \leq 0.33$ & $0.25 \leq \Omega_{m0} \leq 0.36$\\
     & & & $\Omega_{k0} = -0.11$ & $-0.36 \leq \Omega_{k0} \leq 0.06$ & $-0.59 \leq \Omega_{k0} \leq 0.19$\\
     \hline \vspace{2pt}
      & $\phi$CDM & $-0.5 \leq \Omega_{k0} \leq 0.5$ & $\Omega_{m0} = 0.31$ & $0.29 \leq \Omega_{m0} \leq 0.33$ & $0.27 \leq \Omega_{m0} \leq 0.35$\\
     &  & & $\alpha = 1.12$ & $0.51 \leq \alpha \leq 1.59$ & $0.11 \leq \alpha \leq 1.97$\\ 
     \hline \vspace{2pt}
    & & $0.01 \leq \Omega_{m0} \leq 1$ & $\Omega_{k0} = -0.22$ & $-0.38 \leq \Omega_{k0} \leq -0.07$ & $-0.48 \leq \Omega_{k0} \leq 0.03$\\
     &  & & $\alpha = 1.16$ & $0.53 \leq \alpha \leq 1.61$ & $0.12 \leq \alpha \leq 2.01$\\ 
    \hline \vspace{2pt}
    & & $0.01 \leq \alpha \leq 5$ & $\Omega_{m0} = 0.31$ & $0.29 \leq \Omega_{m0} \leq 0.33$ & $0.27 \leq \Omega_{m0} \leq 0.35$\\
     &  & & $\Omega_{k0} = -0.21$ & $-0.38 \leq \Omega_{k0} \leq -0.07$ & $-0.48 \leq \Omega_{k0} \leq 0.04$\\ 
     \hline \hline \vspace{2pt}
     $73.24\pm 1.74$ & XCDM & $-0.7 \leq \Omega_{k0} \leq 0.7$ & $\Omega_{m0} = 0.32$ & $0.29 \leq \Omega_{m0} \leq 0.34$ & $0.26 \leq \Omega_{m0} \leq 0.36$\\
     & & & $w_X = -0.82$ & $-1.08 \leq w_X \leq -0.71$ & $-1.44 \leq w_X \leq -0.63$\\
     \hline \vspace{2pt}
      &  & $0.01 \leq \Omega_{m0} \leq 1$ & $\Omega_{k0} = -0.11$ & $-0.33 \leq \Omega_{k0} \leq 0.03$ & $-0.55 \leq \Omega_{k0} \leq 0.14$\\
     & & & $w_X = -0.82$ & $-1.08 \leq w_X \leq -0.71$ & $-1.44 \leq w_X \leq -0.63$\\
     \hline \vspace{2pt}
     &  & $-2 \leq w_X \leq 0$ & $\Omega_{m0} = 0.32$ & $0.29 \leq \Omega_{m0} \leq 0.34$ & $0.26 \leq \Omega_{m0} \leq 0.36$\\
     & & & $\Omega_{k0} = -0.11$ & $-0.33 \leq \Omega_{k0} \leq 0.03$ & $-0.55 \leq \Omega_{k0} \leq 0.14$\\
     \hline \vspace{2pt}
      & $\phi$CDM & $-0.5 \leq \Omega_{k0} \leq 0.5$ & $\Omega_{m0} = 0.32$ & $0.30 \leq \Omega_{m0} \leq 0.34$ & $0.29 \leq \Omega_{m0} \leq 0.35$\\
     &  & & $\alpha = 0.76$ & $0.31 \leq \alpha \leq 1.14$ & $0.06 \leq \alpha \leq 1.41$\\ 
     \hline \vspace{2pt}
    & & $0.01 \leq \Omega_{m0} \leq 1$ & $\Omega_{k0} = -0.25$ & $-0.40 \leq \Omega_{k0} \leq -0.14$ & $-0.48 \leq \Omega_{k0} \leq -0.07$\\
     & & & $\alpha = 0.79$ & $0.32 \leq \alpha \leq 1.16$ & $0.06 \leq \alpha \leq 1.43$\\ 
    \hline \vspace{2pt}
    & & $0.01 \leq \alpha \leq 5$ & $\Omega_{m0} = 0.32$ & $0.30 \leq \Omega_{m0} \leq 0.34$ & $0.29 \leq \Omega_{m0} \leq 0.35$\\
     &  & & $\Omega_{k0} = -0.24$ & $-0.40 \leq \Omega_{k0} \leq -0.14$ & $-0.48 \leq \Omega_{k0} \leq -0.07$\\ 
     \hline \vspace{2pt}
  \end{tabular}
 \end{table*}

The consensus model, spatially flat $\Lambda$CDM, is consistent with current $H(z)$ + BAO data, but these data allow some nonzero spatial curvature. In particular, we find that the best-fit values of the parameters in the $\Lambda$CDM model imply a curvature energy density parameter of $\Omega_{k0} = 0.03$ for the $\bar{H}_0 \pm \sigma_{H_0} = 68 \pm 2.8$ km s$^{-1}$ Mpc$^{-1}$ prior case, and $\Omega_{k0} = -0.07$ for the $\bar{H}_0 \pm \sigma_{H_0} = 73.24 \pm 1.74$ km s$^{-1}$ Mpc$^{-1}$ prior case. More precisely, using the $\Omega_{m0}$ and $\Omega_{\Lambda}$ best-fit values and error bars for flat \lcdm from Table \ref{table 5}, and combining the errors in quadrature, an approximate estimate is $\Omega_{k0} = 0.03(1 \pm 1.8)$ and $\Omega_{k0} = -0.07(1 \pm 0.59)$ for the \hublow and \hubhigh priors, with the data favoring a closed model at a little over 1$\sigma$ in the second case. The 3-parameter models, in both prior cases, favor closed spatial hypersurfaces, but the error bars are so large that these results only stand out in the \hubhigh prior case of the \pcdm model (see the center and right panels of \ref{fig. 3}). While not very statistically significant, we note that these results are not inconsistent with those of \cite{50, 51, 52} and \cite{38, 53}, who found that CMB anisotropy data, in conjunction with other cosmological data, were not inconsistent with mildly closed spatial hypersurfaces.

The current data are also not inconsistent with some mild dark energy dynamics, although the size of the effect varies depending on the choice of $H_0$ prior and whether or not $\Omega_{k0}$ is allowed to vary as a free parameter. In the flat \pcdm model, for instance, $\alpha$ can be different from zero only in the $\bar{H}_0 \pm \sigma_{H_0} = 68 \pm 2.8$ km s$^{-1}$ Mpc$^{-1}$ prior case, whereas $\alpha$ can be different from zero in both prior cases if $\Omega_{k0}$ is allowed to vary (see the right panel of \ref{fig. 1} and the left panel of \ref{fig. 3}).

\section{Conclusions}
\label{sec. 6}
We analyzed a total of 42 measurements, 31 of which consisted of uncorrelated $H(z)$ data points, with the remainder coming from BAO observations (some correlated, some not), to constrain dark energy dynamics and spatial curvature, by determining how well these measurements can be described by three common models of dark energy: $\Lambda$CDM, the XCDM parametrization, and $\phi$CDM. 

The consensus flat \lcdm model is in reasonable accord with these data, but depending on the model analyzed and the $H_0$ prior used, it can be a little more than 1$\sigma$ away from the best-fit model. These data are consistent with mild dark energy dynamics as well as non-flat spatial hypersurfaces. While these results are interesting and encouraging, more and better data are needed before we can make definitive statements about the spatial curvature of the universe and about dark energy dynamics.

\section*{Acknowledgements}
\label{Ack}
% Entry for the table of contents, for this guide only
\addcontentsline{toc}{section}{Acknowledgements}

The authors thank Narayan Khadka for pointing out a computational error in an earlier version of the paper, thank Omer Farooq and Lado Samushia for useful discussions, and thank Kumar Ayush, Rahul Govind, and Tyler Mitchell for help with computational resources and optimization. This work was supported by DOE Grant DE-SC0011840.

\bibliographystyle{mnras}
\bibliography{bibliography}  %if your bibtex file is called example.bib

% Don't change these lines
\bsp	% typesetting comment
\label{lastpage}
\end{document}